\documentclass[a4paper,article,twocolumn,english,twocolumns,epl]{revtex4}
\usepackage{amsmath}
\usepackage{amsfonts}
\usepackage{amssymb}
\usepackage{comment}
\usepackage{color}
\usepackage{graphicx}
\usepackage{color}
\definecolor{darkblue}{rgb}{0,0,0.6}
\definecolor{darkred}{rgb}{0.6,0,0}
\definecolor{darkgrey}{rgb}{0.6,0.6,0.6}
\usepackage[colorlinks=true,urlcolor=darkblue,citecolor=darkblue,linkcolor=darkred,hyperfootnotes=false]{hyperref}

\providecommand{\U}[1]{\protect\rule{.1in}{.1in}}
\newcommand{\cc}{\text{cr}}
\newcommand{\tr}{\text{tr}}
\newcommand{\LL}{\text{L}}
\newcommand{\RR}{\text{R}}
\newcommand{\ff}{\text{f}}

\newcommand{\bfx}{\mathbf{x}}

\begin{document}
\title{Activated escape of a self-propelled particle from a metastable state}
\author{E. Woillez$^{1}$, Y. Zhao$^{2}$, Y. Kafri$^{1}$, V. Lecomte$^{3}$, J. Tailleur$^{4}$}
\affiliation{$^1$Department of Physics, Technion, Haifa 32000, Israel}
\affiliation{$^2$School of Physics and Astronomy and Institute of Natural Sciences, Shanghai Jiao Tong University, Shanghai, China}
\affiliation{$^3$Universit\'e Grenoble Alpes, CNRS, LIPhy, F-38000 Grenoble, France}
\affiliation{$^4$Universit\'e Paris Diderot, Sorbonne Paris Cit\'e, MSC, UMR 7057 CNRS, 75205 Paris, France}

\begin{abstract} 
We study the noise-driven escape of active Brownian particles (ABPs) and run-and-tumble particles (RTPs) from confining potentials. In the small noise limit, we provide an exact expression for the escape rate in term of a variational problem in any dimension. For RTPs in one dimension, we obtain an explicit solution, including the first sub-leading correction. In two dimensions we solve the escape from a quadratic well for both RTPs and ABPs. In contrast to the equilibrium problem we find that the escape rate depends explicitly on the full shape of the potential barrier, and not only on its height. This leads to a host of unusual behaviors. For example, when a particle is trapped between two barriers it may preferentially escape over the higher one. Moreover, as the self-propulsion speed is varied, the escape route may discontinuously switch from one barrier to the other, leading to a dynamical phase transition.
\end{abstract}

\pacs{
}

\maketitle

Activated escapes from metastable states play a major role in a host
of physical phenomena, with applications in fields as diverse as
biology, chemistry, and
astrophysics~\cite{kampen_stochastic_2007,chandrasekhar1943RMP}. They
also play an important role in active matter, where they control
nucleation in motility-induced phase
separation~\cite{cates2015motility}, activated events in glassy
self-propelled-particle
systems~\cite{berthier_non-equilibrium_2013,nandi2018random}, or
escapes through narrow channels~\cite{paoluzzi_self-sustained_2015}.
However, despite recent
progress~\cite{geiseler2016kramers,demaerel2018active}, little is
known about the physics that controls the rare events leading to the
escape of an active system from a metastable state.

In equilibrium, most of our intuition regarding such events is based
on Kramers seminal work~\cite{kramers1940brownian} on Brownian
particles (see~\cite{hanggi1990reaction} for a review). When the
thermal energy is much lower than the potential barriers, there is
a time-scale separation between rapid equilibration within metastable
states and rare noise-induced transitions between them, a simple
physical picture which is at the root of the modern view on
metastability~\cite{gaveau_theory_1998,bovier_metastability_2015}.  In
this limit, the mean escape time over a potential barrier of height
$\Delta V$ is given by $\left\langle \tau\right\rangle
{\sim}\exp(\frac{\Delta V}{k_{\text{B}}T})$. \if{Note that since the problem is
  governed by rare events there is only a single time scale $\tau$
  appearing in the problem and the activated events are exponentially
  distributed.}\fi At the exponential level, the crossing time over a
potential barrier only depends on its height.

\begin{figure}
\includegraphics[height=3.1cm]{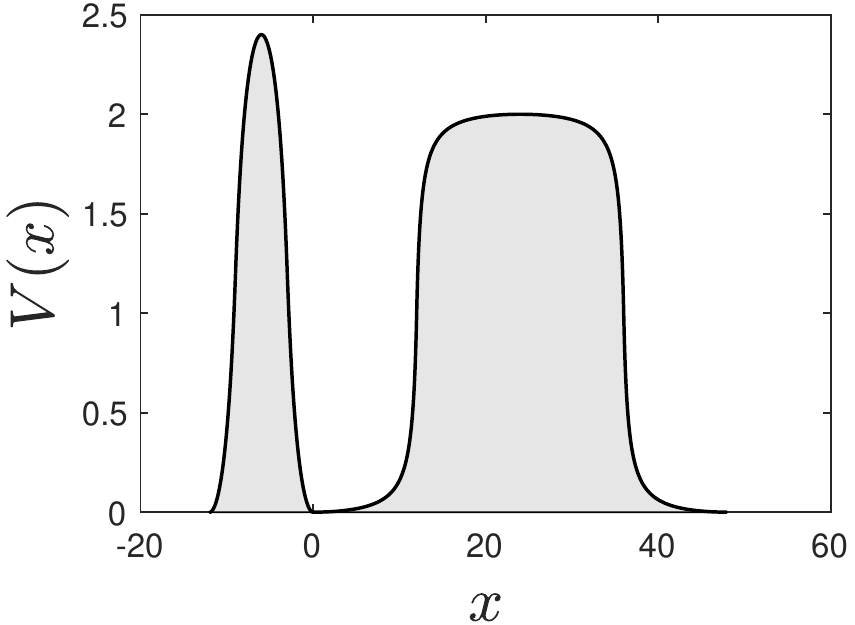}
\includegraphics[height=3.1cm]{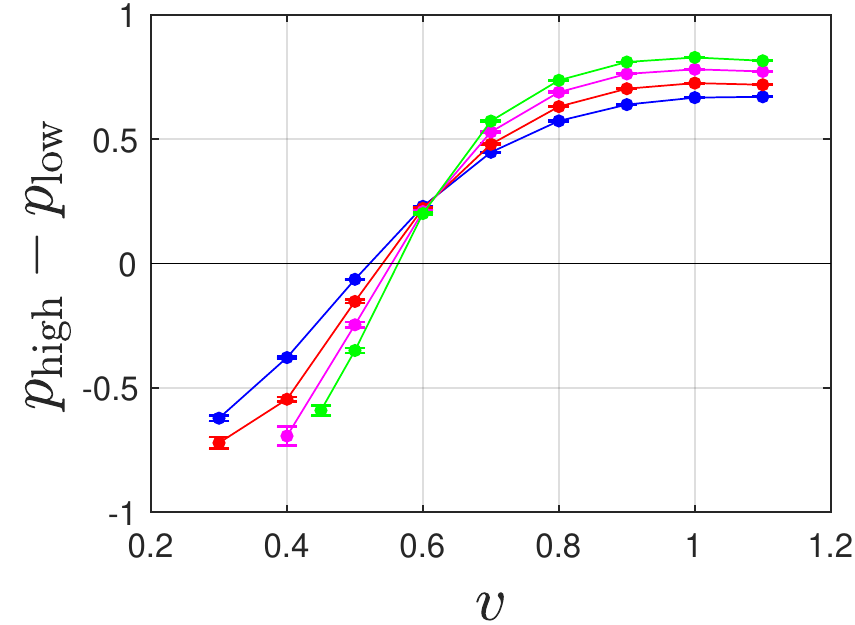}
\caption{Active escape from a metastable well confined by two barriers of different heights ({\bf left}). We measured the fraction of particles escaping over the higher barrier, $p_{\rm high}$, and over the lower one, $p_{\rm low}$, depending on the value of $v$ and with decreasing values of $D$, of Eq.\:(\ref{eq:langevin1D}). The right panel shows that, as $v$ increases, the most likely escape route switches
from the lower barrier to the higher one. The switch between preferred barriers is manifested as a dynamical phase transition in the small-noise limit. This can be seen as the transition becomes sharper when $D$ is decreased (blue: $D=0.08$, red: $D=0.0675$, magenta: $D=0.058$, green: $D=0.051$, colors online).  Details of the potential are given in~\cite{supp}.
\label{fig:twowells}}
\end{figure}

To develop a corresponding intuition for activated processes in active
matter, we follow Kramers and consider the dynamics of an active
particle confined in a metastable well described by a potential $V$:
\begin{equation}
\dot{{\bf x}}=-\mu\nabla V+ v {\bf u}(\theta) +\sqrt{2 D}{\boldsymbol
  \xi}(t).\label{eq:langevin1D}
\end{equation}
Here, ${\bf x}$ is the position of the particle, $v$ its
self-propulsion speed, and $\mu$ its mobility. The orientation of the
particle ${\bf u}(\theta)$ evolves stochastically with a persistence
time $1/\alpha$. Here, $\theta$ is a generalized angle parametrizing the $d-1$ dimensional unit sphere. Finally, ${\boldsymbol \xi}(t)$ is a Gaussian white
noise which may stem from either thermal fluctuations, in which case
$D=\mu\, k_{\text{B}}T$, or from fluctuations of the activity. As we
show below, the escape of such an active particle from a metastable
state is very different from the equilibrium case, leading to a host
of interesting phenomena.  For example, direct simulations of
Eq.\:\eqref{eq:langevin1D} show that active particles confined between
two barriers may preferentially escape over the \textit{higher} one,
depending on the self-propulsion $v$ (See Fig.\:\ref{fig:twowells}).

In what follows, we provide a complete solution of the Kramers problem
for active particles described by Eq.\:\eqref{eq:langevin1D}, in any
dimension, using a path-integral formalism.  In contrast to existing
works on first-passage times~\cite{angelani2017confined,dhar2018run,caprini2019active},
we focus on cases in which the potential is strictly confining at
$D=0$ and the barrier can only be crossed using fluctuations. We refer
to such case as \textit{confining potentials}.  We give an explicit
expression for the mean escape time in terms of a variational problem
for run-and-tumble particles
(RTPs)~\cite{berg_chemotaxis_1972,schnitzer_theory_1993} and active
Brownian particles (ABPs)~\cite{schimansky-geier_structure_1995}, the
latter being studied only in $d\geq 2$ dimensions.
In one dimension, RTPs had previously been studied in the limits
$\alpha\to 0$ and $\alpha \to \infty$~\cite{geiseler2016kramers};
Here, we provide the full solution of the activation time for RTPs for
all $\alpha$, including its sub-exponential prefactor. In cases with
multiple competing reaction paths, our results provide the selection principle for the most
likely escape route. In particular, we explain the dynamical phase transition observed in Fig.\:\ref{fig:twowells}.
 
\begin{figure}
\includegraphics[width=.8\columnwidth]{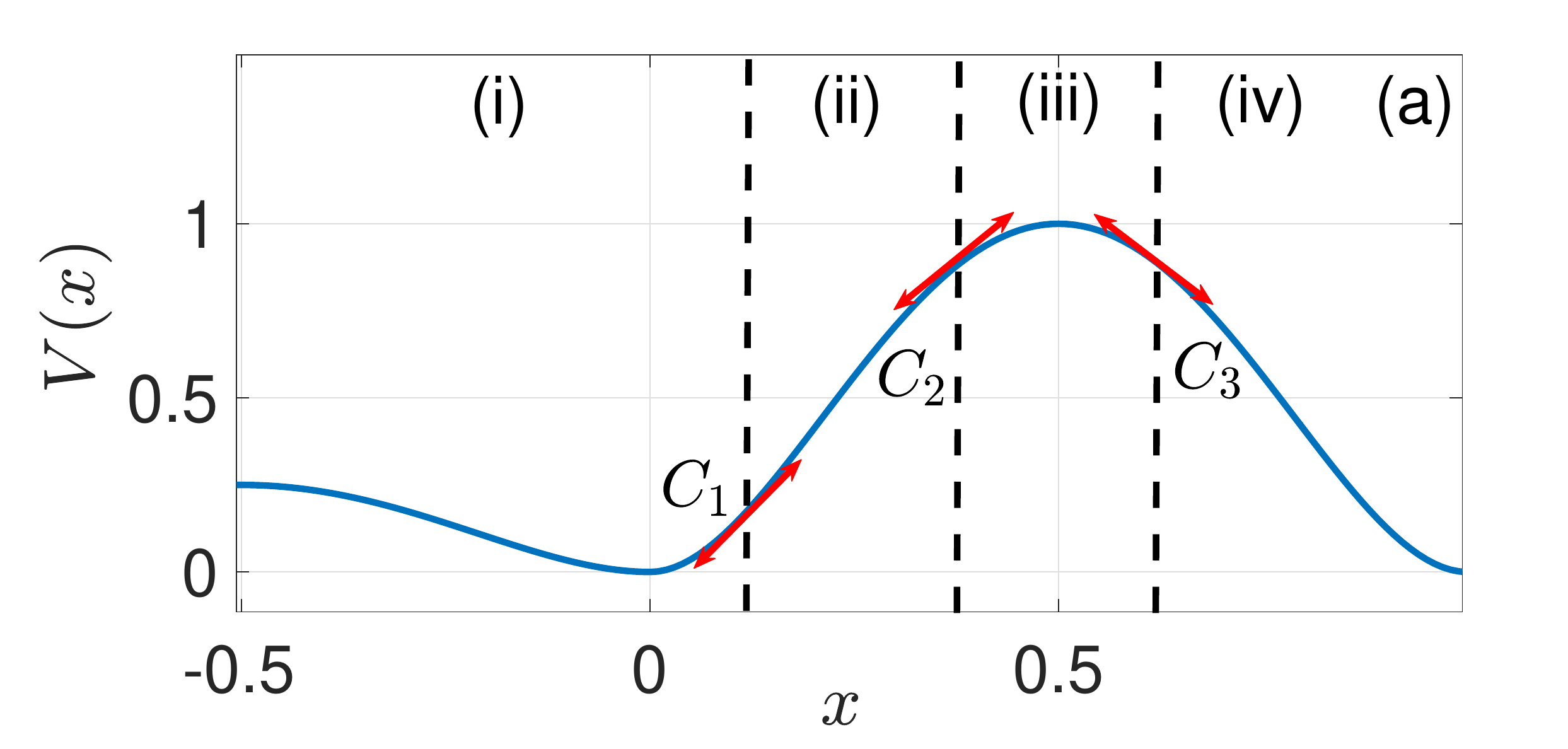}

\hspace*{.5cm}\includegraphics[width=.8\columnwidth]{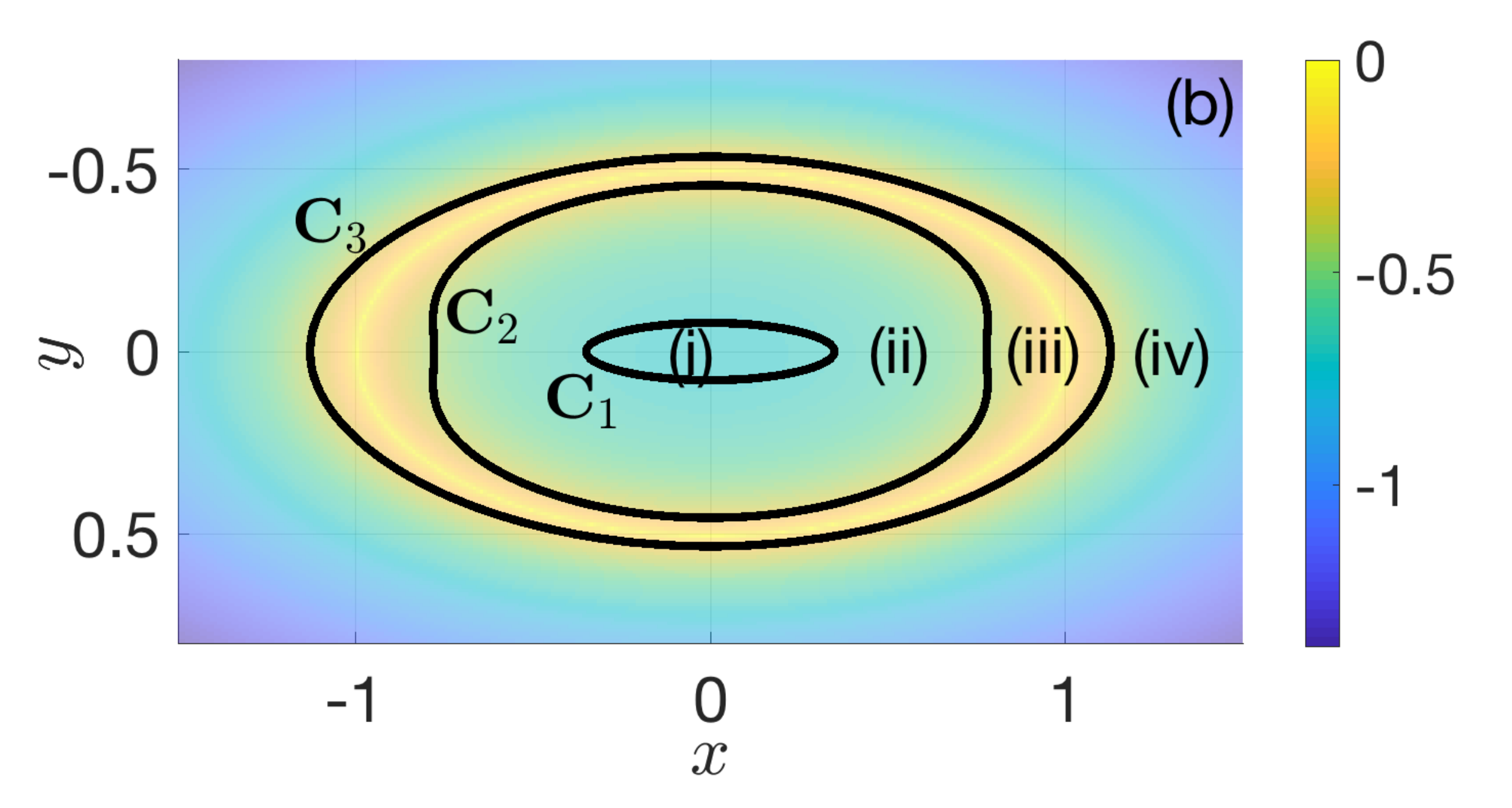}
\caption{Schematic representation of the active escape problem. {\bf
    Top:} Escape in 1d over a barrier, region (i) correspond to the well whereas regions (ii)-(iv) make up the barrier. All are defined in the text. {\bf Bottom:} The color
  code represents the height of the potential. The barrier is located
  around in the yellow region.\label{fig:problem escape}}
\end{figure}

For confining potentials, it is natural to divide the barrier into
separate regions depending on whether the force $|\nabla V|$ is larger
or smaller than the propulsion force $f_p=v/\mu$. Consider, for
instance, the escape in one dimension from a metastable well, see
Fig.\:\ref{fig:problem escape}.  We can identify four different
regions separated by three points $\{C_{1},C_{2},C_{3}\}$ satisfying
$|V'(C_{i})|=f_p$. In regions (i) and (iii), when $x\leq
C_{1}$ or $C_2 \leq x \leq C_3$, the particles feel a force $-V'$
smaller in magnitude than $f_p$. In the $D \to 0$ limit the
contribution of the noise $\xi(t)$ to the dynamics can be
neglected. In region (ii), where $C_1 \leq x \leq C_{2}$, the
particles cannot climb the potential without the noise $\xi(t)$.
Crossing this region is therefore a rare event which controls the
escape from the metastable state. In region (iv), where $x>C_{3}$, the
particles would need the noise to come back to region (i), were they
to reverse direction. This is a rare event and the particle has thus
effectively crossed the barrier once it has reached $C_{3}$. \if{Note
  that when $v=0$, $C_1$ is at the minimum of the metastable potential
  and $C_2=C_3$ is at the maximum of the potential.}\fi The
generalization of these points to lines or surfaces in higher
dimensions (denoted ${\bf C}_i$) is straightforward and an example is
displayed in Fig.\:\ref{fig:problem escape}~\footnote{Note that in
  case of saddles the surfaces ${\bf C}_2$ and ${\bf C}_3$ may merge
  into a single surface with two distinct faces.}. Note that the
problem is activated only if region (ii) exists. Otherwise, the
problem, as considered for example in 1d in
\cite{angelani2014first}, is a first-passage problem with no instanton
physics.
The activated process only corresponds to moving across region (ii) so
that the crossing probability is given, to leading order, by 
histories connecting points on ${\bf C}_1$ and ${\bf C_2}$.  To obtain
the escape time we then write the transition probability
$P(\mathbf{x_2},t|\mathbf{x_{1}},0)$ to be at $\mathbf{x_2} \in {\bf
  C}_2$ at time $t$ starting at $\mathbf{x_{1}} \in {\bf C}_1$ as a
path integral in its Onsager--Machlup
form~\cite{onsager_fluctuations_1953}
\begin{equation}
P(\mathbf{x}_2,t|\mathbf{x_{1}},0)=\int_{\bf x_1}^{\bf x_2}\mathcal{D}\left[\mathbf{x}(t),\theta(t)\right]e^{-\frac{1}{D}\mathcal{A}[\mathbf{x},\theta]}{\cal P}[\theta(t)]\label{eq:path integral}\;.
\end{equation}
${\cal P}[\theta(t)]$ is the probability of a history of the angle
$\theta$. For example, ABPs in 2d with rotational diffusivity $\alpha$ lead to
${\cal P}[\theta(t)]\propto e^{-\int_{0}^{t}
  \dot{\theta}^{2}/(4\alpha){\rm d}t'}$. In Eq.\:\eqref{eq:path integral},
the action $\mathcal{A}[\mathbf{x},\theta]$ is given by
\begin{equation}
\mathcal{A}[\mathbf{x},\theta]=\frac{1}{4}\int_{0}^{t}\left\Vert \mathbf{\dot{x}}+\mu\nabla V(\mathbf{x})-v\mathbf{u}(\theta)\right\Vert ^{2}{\rm d}t'.\label{eq:total action}
\end{equation}
We first integrate expression (\ref{eq:path integral}) over the paths
$\theta(t)$ to obtain an effective action for the probability of a
path $\mathbf{x}(t)$. In the limit $D\rightarrow0$, we use a
saddle-point approximation in (\ref{eq:path integral}) to get:
\begin{equation}
\int\mathcal{D}[\theta(t)]e^{-\frac{1}{D}\mathcal{A}[\mathbf{x},\theta]}{\cal P}[\theta(t)] \underset{D\rightarrow0}{\asymp}               e^{-\frac{1}{D}\mathcal{A}[\mathbf{x},\tilde{\theta}]}\label{eq:space path LDP}
\end{equation}
where $\asymp$ stands for logarithmic equivalence and
$\tilde{\theta}(t)$ is the path satisfying the variational problem
\begin{equation}\label{eq:action27}
\mathcal{A}[\mathbf{x},\tilde{\theta}]=\underset{\theta}{\inf}\left\{ \frac{1}{4}\int_{0}^{t}\left\Vert \mathbf{\dot{x}}+\mu\nabla V(\mathbf{x})-v\mathbf{u}(\theta)\right\Vert ^{2}{\rm d}t'\right\} .
\end{equation}
Note that ${\cal P}[\theta(t)]$ is a subdominant contribution and any
cost to the action arising from it can be ignored to leading
order \footnote{The results might change in cases where, say for ABPs,
  $\alpha$ is proportional to $D$.}.  Clearly, the optimum requires
$\mathbf{u}(\theta)$ to be in the same direction as
$\mathbf{\dot{x}}+\mu\nabla V(\mathbf{x})$ so that
\begin{equation}
\mathbf{u}(\tilde{\theta})=\frac{\mathbf{\dot{x}}+\mu \nabla V(\mathbf{x})}{\left\Vert \mathbf{\dot{x}}+\mu\nabla V(\mathbf{x})\right\Vert }\;.\label{eq:optimal angle}
\end{equation}
Using Eqs.\:(\ref{eq:optimal angle}) and (\ref{eq:total action}), we
find that the transition probability between ${\bf x}_1$ and ${\bf
  x}_2$ is dominated by paths which minimize the action
\begin{equation}
\mathcal{A}[\mathbf{x}]=\frac{1}{4}\int_{-\infty}^{\infty}\left(\left\Vert \mathbf{\dot{x}}+\mu\nabla V(\mathbf{x})\right\Vert -v\right)^{2}{\rm d}t'\;,\label{eq:path action}
\end{equation}
where we have sent the limits of the integral to $\pm\infty$, using
the fact that extremal trajectories start and end at stationary points
(see for
instance~\cite{Tailleur2008JPA,baek2015singularities}). Finally, the
escape time is given by
\begin{equation}
\begin{cases}
\left\langle \tau\right\rangle& \underset{D\rightarrow0}{\asymp}e^{\frac{\phi}{D}}\;,\\
\phi& =\underset{\lbrace \mathbf{x}_1\in {\bf C}_1,\mathbf{x}_2\in {\bf C}_2\rbrace}{\inf}\: \underset{{\bf x}(t)}{\inf} \mathcal{A}[\mathbf{x}(t)] \;.
\end{cases}\label{eq:quasipot2D}
\end{equation}
The inner minimization corresponds to optimizing the action over
different paths; it is realized by an instanton $\mathbf{x}(t)$ which
connects $\mathbf{x}_1$ and $\mathbf{x}_2$. The outer minimization
corresponds to optimizing over all possible initial and final
positions of the instanton. Eq.\:\eqref{eq:quasipot2D} provides a full
solution to the escape problem for both ABPs and RTPs as a variational
problem. It generalizes the Kramers law and we discuss the physics of
the quasi-potential barrier $\phi$ below.  Note that when $v=0$ the minimizers
of the action are $\mathbf{\dot{x}}=\mu \nabla V(\mathbf{x})$ and we
recover the usual Kramers law with $\phi=\mu \,\Delta V$, where
$\Delta V$ is the minimal potential difference across the barrier.
We now turn to apply our results to a general one-dimensional potential
barrier and to an elliptic well in two dimensions.

\begin{figure}
    \centering
    \includegraphics[width=.49\columnwidth]{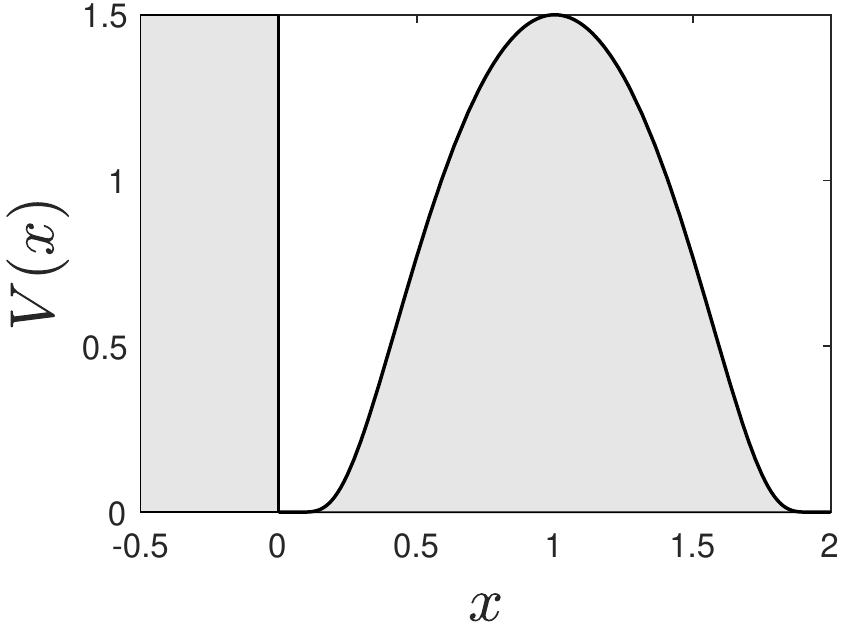}
    \includegraphics[width=.49\columnwidth]{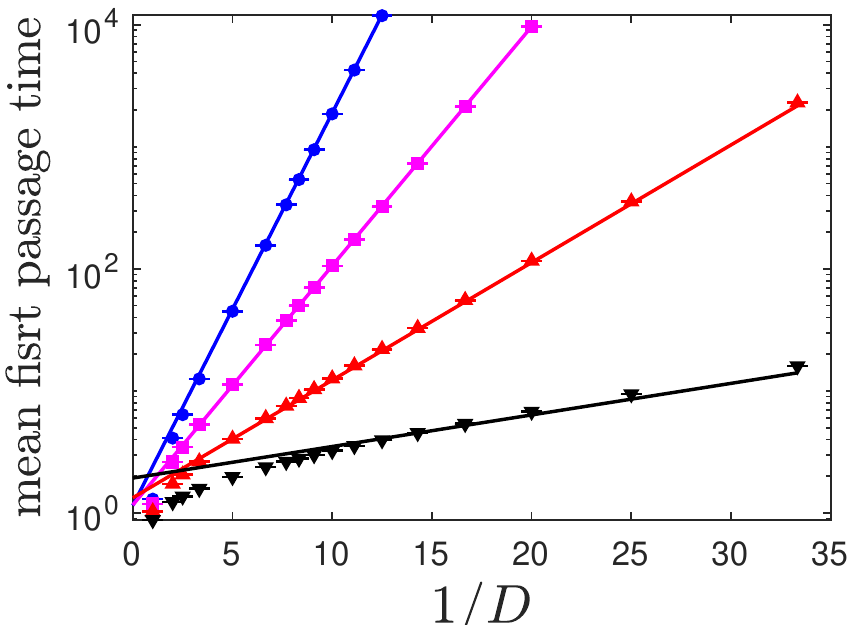}
    \caption{We compute the mean-first passage time $\langle\tau\rangle$ over a confining barrier. Details of the potential shown in the left panel are given in~\cite{supp}. The right panel shows the validity of our generalized Kramers law for several values of $v= 1.0, 1.5, 2.0, 2.5$}.
    \label{fig:checkcheck}
\end{figure}

\noindent {\it RTPs in one dimension:} Here, ${\bf u}(\theta)$ is replaced by a binary variable $u=\pm
1$ which flips with rate $\alpha/2$. As in Fig.\:\ref{fig:problem
  escape}, the barrier is located on the right of the metastable well.
$\bfx_{1,2}$ are then given by $C_{1,2}$. Clearly, the minimal action
is obtained by particles with $u=1$: particles which reverse their
motion in the middle of the instanton are exponentially less likely to
cross the barrier. The action then reduces to
\begin{equation}
\mathcal{A}[x]=\frac{1}{4}\int_{-\infty}^{\infty}\left[ \dot{x}+\mu V'(x) -v\right]^{2}{\rm d}t'.\label{eq:path action_1d}
\end{equation}
It is thus equivalent to an equilibrium problem in an effective titled
potential $\varphi(x)/\mu$; the instanton solution  obeys
\begin{equation}
 \dot{x}=  \partial_x \big[ \mu\,(V(x)-V(C_1))-v\,(x-C_1)	 \big]\equiv \partial_x \varphi(x)\;, \label{eq:varphi}
\end{equation}
which gives, for the quasi-potential barrier introduced in Eq.\:(\ref{eq:quasipot2D}),
\begin{equation}
  \phi=\mu \left[V(C_2)-V(C_{1})\right]-v\left(C_2-C_{1}\right)\;.\label{eq:quasipot}
\end{equation}
Our predictions~\eqref{eq:quasipot2D} and~\eqref{eq:quasipot} are verified in Fig.~\ref{fig:checkcheck} using direct simulation of Eq.~\eqref{eq:langevin1D} with a single barrier. 

Using asymptotic techniques~\cite{supp,bouchet2016generalisation}, we also obtain the
leading sub-exponential amplitude of the transition
time~\eqref{eq:quasipot2D}. For simplicity we consider a boundary
condition in which the potential is flat on the left of the barrier
and the density of particles in that region is $\rho_0$; other
boundary conditions are discussed in~\cite{supp}. The mean time between
particles crossing the barrier is then given by $\langle \tau \rangle
\underset{D\rightarrow0}{\sim} A e^{\frac{\phi}{D}}$, where
%\begin{widetext}
%\begin{equation}
%A=\frac{2\pi e^{-\frac \alpha 2 {\mathcal T}_{\rm inst}}}{\rho_{0}v^{2} \Gamma(1-\frac{\alpha }{ 2k_2})\Gamma(\frac{\alpha }{ 2k_1})}\frac{\left[\frac{D}{v^2} k_1\right]^{\frac{k_1-\alpha}{2k_1}}}{\left[\frac{D}{v^2}|k_2|\right]^{\frac{k_2-\alpha}{2 k_2}}}\frac{\int_{C_{2}}^{C_{3}}\left(\alpha-\mu V''(y)\right)e^{\alpha\mathfrak{F} \int_{C_{2}}^{y}\frac{\mu V'(z)}{v^{2}-\left(\mu V'(z)\right)^{2}}{\rm d}z }{\rm d}y}{e^{-\alpha\mathfrak{F} \int_{-\infty}^{C_{1}}\frac{\mu V'(y)}{v^{2}-\left(\mu V'(y)\right)^{2}}{\rm d}y }} \if{e^{\frac{\alpha}{2}\mathfrak{F}\left\{ \int_{C_{1}}^{C_{2}}\frac{{\rm d}y}{\varphi'(y)}\right\}} }\fi.\label{eq:big formule 1D}
%\end{equation}
%\end{widetext}
\begin{eqnarray}
A&=&\frac{2\pi e^{-\frac \alpha 2 {\mathcal T}_{\rm inst}}}{\rho_{0}v^{2} \Gamma(1-\frac{\alpha }{ 2k_2})\Gamma(\frac{\alpha }{ 2k_1})}\frac{\left[\frac{D}{v^2} k_1\right]^{\frac{k_1-\alpha}{2k_1}}}{\left[\frac{D}{v^2}|k_2|\right]^{\frac{k_2-\alpha}{2 k_2}}} \\
&&\times\frac{\int_{C_{2}}^{C_{3}}\left(\alpha-\mu V''(y)\right)e^{\alpha\mathfrak{F} \int_{C_{2}}^{y}\frac{\mu V'(z)}{v^{2}-\left(\mu V'(z)\right)^{2}}{\rm d}z }{\rm d}y}{e^{-\alpha\mathfrak{F} \int_{-\infty}^{C_{1}}\frac{\mu V'(y)}{v^{2}-\left(\mu V'(y)\right)^{2}}{\rm d}y }}.\nonumber\label{eq:big formule 1D}
\end{eqnarray}
Here, ${\cal T}_{\rm inst}=\mathfrak{F} \int_{C_{1}}^{C_{2}}\frac{{\rm d}y}{\partial_{y}\varphi}$ is the duration of the instanton,
$k_i=\mu V''(C_i)$, $\Gamma(x)$ is the Euler Gamma function, and
$\mathfrak{F}$ denotes the finite part of the integral, defined by
removing the logarithmic divergences occurring at $C_1$ and $C_2$,
e.g.
\begin{equation}
  \begin{aligned}
    {\mathfrak{F}} \int_{-\infty}^{C_{1}}\frac{\mu V'(y)}{v^{2}-\left(\mu V'(y)\right)^{2}}{\rm d}y &  =\\ 
\underset{x \to C_1}\lim \Big\lbrace \int_{-\infty}^{x}\frac{\mu V'(y)}{v^{2}-\left(\mu V'(y)\right)^{2}}{\rm d}y & + \frac{1}{2k_1}\log\left(\frac{k_{1}(C_{1}-x)}{v}\right) \Big\rbrace \;.
  \end{aligned}\label{eq:prefactor}
\end{equation}
The term $e^{-\frac{\alpha}{2}{\cal T}_{\rm inst} }$ has a simple
interpretation: it is the probability that the particle does not flip
along the instanton. Note that the $v=0$ limit is singular: all
histories of  $u(t)$ are then equally likely, a
degeneracy which otherwise does not exist.
\if{The other terms are the equivalent of the 
$1/\sqrt{\left|\partial_x^2 V(C_2)\right| \partial_x^2 V(C_1)}$ 
term appearing in the $v=0$ problem. Interestingly, when $v \neq 0$ they depend on the exact shape of the potential in the regions $x<C_1$ and $C_2< x < C_3$. Nonetheless, second derivatives of the potential, but now at the points $C_1$ and $C_2$, and encoded in $k_1$ and $k_2$, appear in the expression. Note that the lifted degeneracy due to the finite value $v>0$ manifests itself in the solution, for example, through the terms accounting for the time of the instanton.}\fi

Equations\:\eqref{eq:quasipot} and~\eqref{eq:prefactor} provide an explicit solution to the Kramers problem in one dimension.  Note
that the effect of the activity cannot be cast into a simple
description with an effective temperature. Both $\phi$ and the
prefactor indeed depend on the full functional form of the potential
$V$. 

{\noindent\it Dynamical phase transition:} We now show how the
analysis of the quasi-potential accounts for the non-trivial choice of
escape routes when the particle is trapped between two potential
barriers. In the small $D$ limit, the escape time is controlled by the
quasi-potential~\eqref{eq:quasipot} of each barrier, which we can
study separately. For the right barrier, the explicit dependence of
$\phi$ on $v$ reads
\begin{equation}
\phi(v)=\mu\left[V(C_{2}(v))-V(C_{1}(v))\right]-v[C_{2}(v)-C_{1}(v)]\;.
\end{equation}
When $v=0$, we recover the standard Kramers result
$\phi(0)=\mu\left(V(C_2)-V(C_1)\right)$. Using $\mu V'(C_1)=\mu
V'(C_2)=v$, one has $\phi'(v)=-(C_{2}-C_{1})$, which implies that
$\phi$ is a decreasing function of $v$. When $v>
v_{\cc}\equiv\underset{x}{\max}\left\{ \mu| V'(x)|\right\} $, the
particle can cross the barrier without thermal activation so that
$\phi(v_{\cc})=0$. $\phi(v)$ thus decreases from the equilibrium $v=0$
value to zero. The initial decrease of the escape time is given by
$\phi'(0)=-[C_{2}(0)-C_{1}(0)]\equiv -\ell$ which is nothing but the
distance between the maxima and the minima of the potential $V$, i.e. the
width of the barrier. The same construction holds for the second
barrier.

\begin{figure}
\includegraphics[height=4cm]{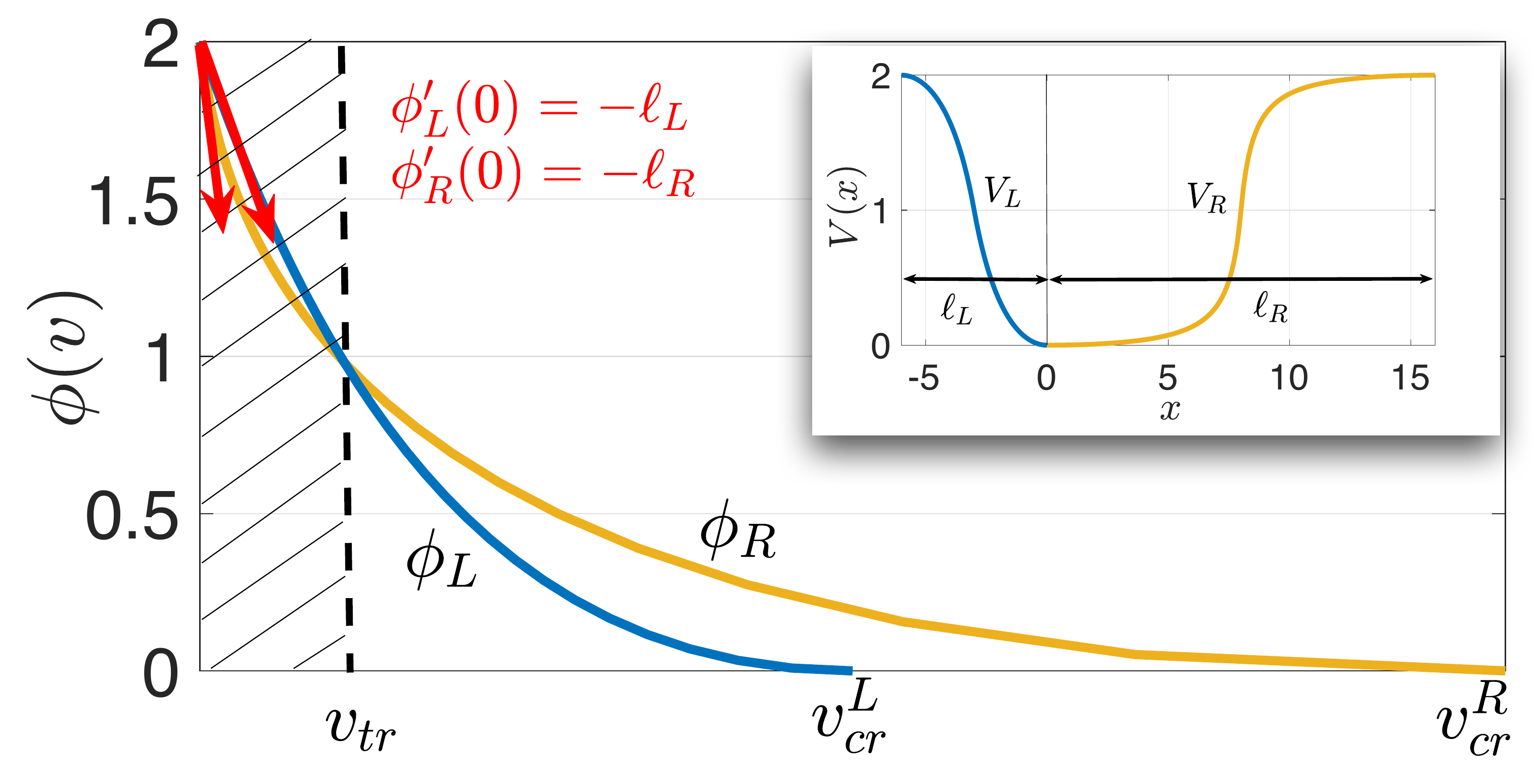}
\caption{The first panel displays the trap with two asymmetric escape walls
$V_{\LL}$ and $V_{\RR}$. The second panel displays the two quasi-potentials
$\phi_{\LL}(v)$ and $\phi_{\RR}(v)$ as functions of $v$ (further explanations
in the text). This illustrates the dynamical phase transition where
$v$ is the control parameter. For $v=0$, particles have the same
probability of escape (at the exponential level) through both sides.
For $0<v<v_{\tr}$ (hatched area), particles escape to the right, and for $v_{\tr}<v$
they escape to the left. \label{fig:transition escape}}
\end{figure}

Next, consider the two potential barriers $V_{\RR,\LL}(x)$ of equal
height described in Fig.\:\ref{fig:transition escape}. The right
barrier is wider, $\ell_\RR>\ell_\LL$, but has a larger maximal slope
than the left barrier so that $v_\cc^\RR > v_\cc^\LL$. To leading
order, the escape rates over the two barriers for $v=0$,
$\phi_{\LL}(0)$ and $\phi_{\RR}(0)$, are equal. Following the above
discussion, $\phi_\RR(v)$ decreases faster than
$\phi_\LL(v)$ near $v=0$ because the right barrier is wider than the left one:
for small $v$, the particle is more likely to escape over the right
barrier. $\phi_\RR(v)$ however vanishes at a value $v_\cc^\RR$ larger
than $v_\cc^\LL$ due to the existence of a steeper portion in the
right barrier. For large $v$, the escape is thus more likely through
the left barrier. Hence, there exists a critical self-propulsion speed
at which the most likely escape route changes discontinuously. The
physics presented in Fig.\:\ref{fig:twowells} can be understood from
the above discussion, the sole difference being that the escape rates
are different at $v=0$ due to the different barrier heights. In the $D
\to 0$ limit, the sigmoid function presented in Fig\:\ref{fig:twowells} hence converges to a discontinuous step function. In fact,
it is straightforward to see that one could also observe not one but
two successive dynamical phase transitions if the larger and steeper
barrier were also higher. Interestingly, the dependence of the escape
time on $v$ can be used to sort active particles depending on their
velocities (See Supplementary movie).

\begin{figure}
\vspace{0.3cm}
\includegraphics[height=4cm]{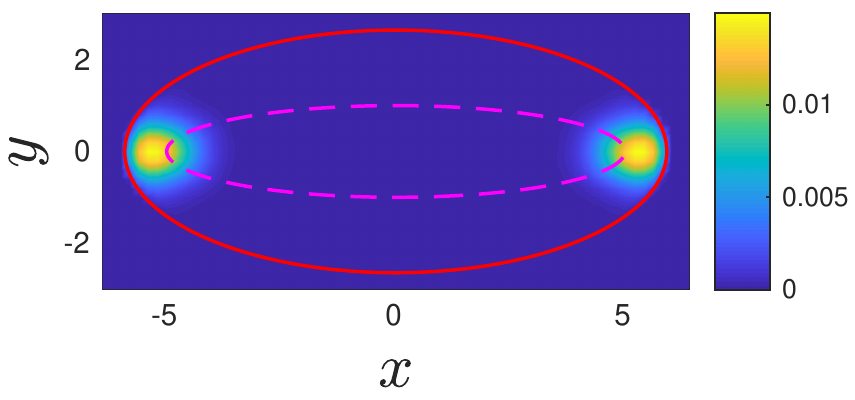}

\hspace*{-1.5cm}\includegraphics[height=4cm]{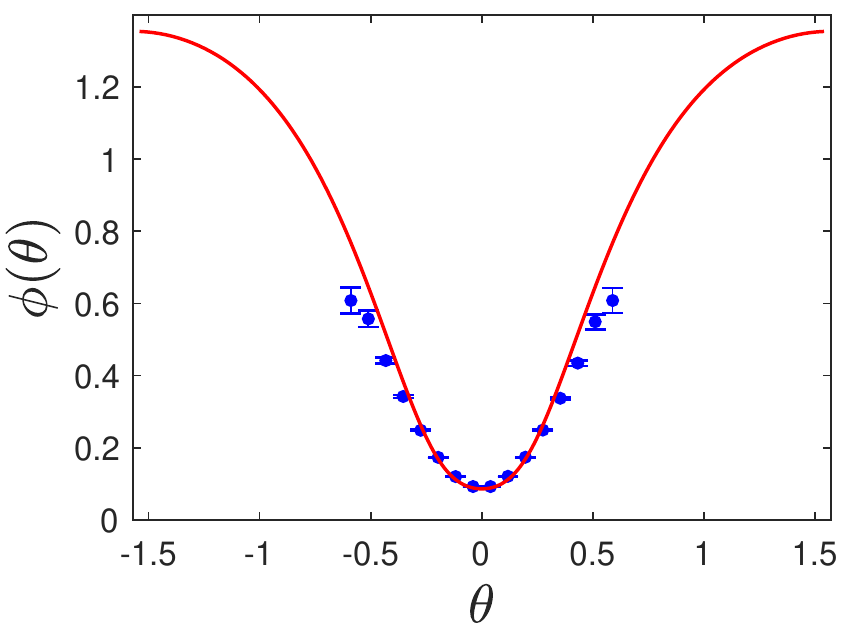}
\caption{Active escape from an elliptic trap ({\bf Top}). Activated escapes have to go from the curve $C_1$ (purple) up to the trap boundary $C$ (red) defined by $V(\mathbf{x})=V_0$. Color encodes the density of particles during the last $\delta t=0.05$ before the escapes, highlighting the preferential route through the apices of the elliptic well. {\bf Bottom:} numerical (dots) and analytical (curve) computation of the quasi-potential $\varphi$ along $C$ (up to a trivial geometric Jacobian) parametrized by $\theta\equiv\arctan(y/x)$. As expected, the quasi-potential reaches a minimum on the major axis (direction $\mathbf{e_{x}}$). For an equilibrium system, the quasi-potential would be flat in the $D\to 0$ limit. \label{fig:2Dwell}}
\end{figure}

\noindent {\it Escape from two-dimensional elliptic potentials:} 
We now consider the escape  of active particles
from a two-dimensional potential well of the form
\begin{equation}
V(x,y)=\lambda_{m}\frac{x^{2}}{2}+\lambda_{M}\frac{y^{2}}{2}\;,\label{eq:elliptic pot}
\end{equation}
with $\lambda_M>\lambda_m$ (for an analysis of the steady-state distribution for the case $\lambda_m=\lambda_M$, see \cite{malakar2019exact}). We assume that particles escape when they
reach a given height $V(x,y)=V_{0}$. This level line $\mathbf{C}$
replaces $\mathbf{C}_2$ of the general discussion, see
Fig.\:\ref{fig:2Dwell}. 

The most-probable escape routes can be computed by solving the
Euler-Lagrange equations for the action given in Eq.\:(\ref{eq:path
  action}), as detailed in the SI. Following the previous
argument we introduce
\begin{equation}
\varphi(\mathbf{x}_{\ff}) \equiv\underset{\mathbf{x(t)}}{\inf}\big\{ \mathcal{A}[\mathbf{x}(t)]\:\big|\:\mathbf{x}(-\infty)\in\mathbf{C}_{1},\mathbf{x}(\infty)=\mathbf{x}_{\ff}\big\} .
\label{eq:quasipot ellispe}
\end{equation}
which yields, at the exponential level, the probability to reach any point $\mathbf{x}_{\ff}$ on the boundary. This log-probability, which we compute in~\cite{supp}, is plotted, in Figure\:\ref{fig:2Dwell}, as a function of the angular parametrization of $\mathbf{x}_{\ff}$, and compared with numerics. Interestingly, the quasi-potential is not constant over the boundary: the particles have a much larger
probability to escape in the direction of the major axis of the
ellipse. This is the most striking difference with the equilibrium problem: For passive Brownian particles, the
quasi-potential is $\varphi(\mathbf{x})=\mu V(\mathbf{x})$, so that
particles have an equal probability (at the exponential level) to
escape through any point along the boundary ${\bf C}$. Activity thus breaks the equilibrium quasi-potential symmetry.

Furthermore, one can compute explicitly the full expression of $\phi$ given by the minimum of the function $\varphi(\mathbf{x}_{\ff})$
along $\mathbf{C}$:
\begin{equation}
\phi=\mu V_{0}\left(1-\sqrt{\frac{v^{2}}{2\mu^{2}\lambda_{m}V_{0}}}\right)^{2}.
\end{equation}
The escape time from the elliptical well is then given by
$\langle \tau \rangle \asymp \exp(\phi/D)$. It solely depends on the potential height, the particle speed, and the semi-axis corresponding to the most likely exit direction. As expected, we recover the standard equilibrium result $\phi=\mu V_{0}$ when $v=0$. 

By providing a full solution to the Kramers problem for both ABPs and
RTPs in any dimensions, we have highlighted how the physics of these
non-equilibrium systems is very different from that of the equilibrium
problem. In particular, the activation barrier, encoded in the
quasi-potential, is not solely defined by the height of the potential
well. Instead, it corresponds to the region where the self-propelling
force fails to overcome the confining one, leading to activation paths
and times that depend in a non-trivial way on both the
self-propelling speed and the full shape of the potential, and to a
wealth of unusual features. Our results also highlight why an effective equilibrium approach is inappropriate. Beyond the case addressed here of an external potential, escape problems play an important role in a host of collective phenomena, from nucleation to glassy physics. It will thus be very interesting to see how the phenomena uncovered in this manuscript play a role in these more complicated systems.   

\noindent {\it Acknowledgments:} YK \& EW acknowledge support from I-CORE Program of the Planning and Budgeting Committee of the Israel Science Foundation and an Israel Science Foundation grant. JT is funded by ANR Bactterns. JT \& YK acknowledge support a joint CNRS-MOST grant. VL is supported by the ERC Starting Grant No.\:68075 MALIG, the ANR-18-CE30-0028-01 Grant LABS and the ANR-15-CE40-0020-03 Grant LSD.

\bibliographystyle{plain_url}
\bibliography{biblio}

\appendix
\begin{widetext}

\section{Path integral formulation}
In this section, we give a simple derivation of the path integral
formulation Eq. (2) in the main text. Let $\boldsymbol{\xi}(t)$ be
the standard Gaussian white noise in $d$ dimensions with correlation
function $\left\langle \xi_{i}(t')\xi_{j}(t)\right\rangle =\delta_{ij}\delta(t-t')$.
The probability of a given realization of $\boldsymbol{\xi}(t)$ on
the interval $[0,T]$, denoted $\mathcal{P}[\boldsymbol{\xi}(t)]$,
can be formally written as 
\begin{equation}
\mathcal{P}[\boldsymbol{\xi}(t)]=\mathcal{N}e^{-\frac{1}{2}\int_{0}^{T}\left\Vert \boldsymbol{\xi}(t)\right\Vert ^{2}{\rm d}t},\label{eq:proba Brownian}
\end{equation}
 where $\mathcal{N}$ is the normalization factor. From Eq. (1) in
the main text, it appears that the probability distribution of any
path $\mathcal{P}[\mathbf{x}(t)]$ can be expressed from the distribution
of the noise $\mathcal{P}[\boldsymbol{\xi}(t)]$ through the change
of variable
\begin{equation}
\frac{1}{\sqrt{2D}}\left[\mathbf{\dot{x}}(t)-\left(-\mu\nabla V(\mathbf{x}(t))+v\mathbf{u}(\theta(t))\right)\right]=\boldsymbol{\xi}(t),\label{eq:change variable}
\end{equation}
where $\theta(t)$ is a given realization of the angle history. Combining
(\ref{eq:proba Brownian}-\ref{eq:change variable}), the probability
of a given path $\mathbf{x}(t)$ conditioned on a realization $\theta(t)$
becomes
\[
\mathcal{P}[\left.\mathbf{x}(t)\right|\theta(t)]=\mathcal{N}'e^{-\frac{1}{4D}\int_{0}^{T}\left\Vert \mathbf{\dot{x}}(t)-\left(-\mu\nabla V(\mathbf{x}(t))+v\mathbf{u}(\theta(t))\right)\right\Vert ^{2}{\rm d}t},
\]
 where $\mathcal{N}'$ is the new normalization factor.  Note that in the It\^o convention of stochastic calculus, the Jacobian  of the change of variable  is a constant independent of the field $\mathbf{x}(t)$, hence the new normalization factor $\mathcal N'$. As the dynamics
of $\theta(t)$ is decoupled from that of $\mathbf{x}(t)$, the joint
probability of $[\mathbf{x}(t),\theta(t)]$ is given by $\mathcal{P}[\mathbf{x}(t),\theta(t)]=\mathcal{P}[\left.\mathbf{x}(t)\right|\theta(t)]\mathcal{P}[\theta(t)]$.
Integrating the joint probability $\mathcal{P}[\mathbf{x}(t),\theta(t)]$
over all possible angle histories and all possible paths $\mathbf{x}(t)$
joining $\mathbf{x}_{1}$ to $\mathbf{x}_{2}$ gives the result presented
in Eq. (2) and Eq. (3) in the main text.

\section{Run-and-Tumble particles in one-dimension}

The most straightforward method to obtain the leading order behavior of the escape rate is presented in the main text. However, in order to find
the sub-leading correction it is useful to employ a different approach involving asymptotic matching of solutions. In what follows this approach, whose final results is Eq.\:(12) of the main text, is detailed. In addition to the result of the main text we also provide here the prefactor for the mean escape time from a metastable well in Eq.\:(\ref{eq:metastable}) of Sec.\:\ref{sec_metastable}.

\subsection{Description of the problem and main equations}

We study RTPs particles in one-dimension. The particles experience a driving force $v/\mu$ which reverses its direction with rate $\alpha/2$. In addition, they are subject to an external potential $V$.
Denoting by $P_{+}(x,t)$ and $P_{-}(x,t)$ the probability density of particles moving to the
right and left respectively, the Fokker--Plank equation for $P_{+},P_{-}$
is 
\begin{equation}
\begin{cases}
\partial_{t}P_{+}= & -\partial_{x}\left[\left(v-\mu\partial_{x}V\right)P_{+}\right]-\frac{\alpha}{2}\left(P_{+}-P_{-}\right)+D\partial_{x}^{2}P_{+},\\
\partial_{t}P_{-}= & -\partial_{x}\left[\left(-v-\mu\partial_{x}V\right)P_{-}\right]-\frac{\alpha}{2}\left(P_{-}-P_{+}\right)+D\partial_{x}^{2}P_{-},
\end{cases}\label{eq:nondim swim}
\end{equation}
Here, as in the main text, $\mu$ is the mobility and $D$ is the diffusion coefficient. We are interested in the limit $D\rightarrow0$, which can be interpreted physically as the asymptotic regime $D\ll v\ell$ where $\ell$ is the barrier length.
Let $\rho=P_{+}+P_{-}$ be the total density of active swimmers in
the medium, and $m=P_{+}-P_{-}$. From Eq.\:(\ref{eq:nondim swim})
\begin{equation}
\begin{cases}
\partial_{t}\rho= & -\partial_{x}\left[vm-\rho\mu\partial_{x}V-D\partial_{x}\rho\right],\\
\partial_{t}m= & -\partial_{x}\left[v\rho-v\mu\partial_{x}V\right]-\alpha m+D\partial_{x}^{2}m.
\end{cases}\label{eq:full equations}
\end{equation}
The first equation describes the mass conservation with a flux 
\[
j(x)=vm -\rho\mu\partial_{x}V-D\partial_{x}\rho,
\]
which is constant $j(x)=J$ in the steady-state. This gives
\begin{equation}
vm-\rho\mu\partial_{x}V-D\partial_{x}\rho=J.\label{eq:flux expression}
\end{equation}
Using this relation in the second equation of\:(\ref{eq:full equations}) we have
\begin{equation}
-\partial_{x}\left[v^{2}\rho-\left(\rho\mu\partial_{x}V+D\partial_{x}\rho\right)\mu\partial_{x}V\right]-\alpha\left(\rho\mu\partial_{x}V+D\partial_{x}\rho\right)+D\partial_{x}^{2}\left(\rho\mu\partial_{x}V+D\partial_{x}\rho\right)=\left(\alpha-\mu\partial_{x}^{2}V\right)J.\label{eq:close density}
\end{equation}
We now solve this equation using standard asymptotic matching techniques with the boundary conditions
\begin{equation}
\begin{cases}
\rho(x)\underset{x\rightarrow-\infty}{\longrightarrow}\rho_{0},\\
\rho(C_{3})=0.
\end{cases}
\label{eq:reservoir_BC}
\end{equation}
In this configuration, the transition of particles across the barrier
is a Poisson process with rate $J$. The mean waiting time between
two particles crossing the barrier is given by $\left\langle \tau\right\rangle =\frac{1}{J}$. As stated above, we will also consider the situation where the particles start from a metastable state (instead of the boundary conditions described by Eq.\:(\ref{eq:reservoir_BC})) and provide an explicit expression of the mean escape time in that case.

\subsection{Methods}

To proceed we solve the problem in the three regions (i), (ii), and (iii) defined in the main text and then match the solutions. We first note the following about the different regions.
\begin{enumerate}
\item \textbf{region (i): }The flux $J$ is so small compared to the other terms
in Eq.\:(\ref{eq:close density}) that the solution is given by
the steady-state with $D=0$ and $J=0$. The corrections are of order $D$. Namely, we solve
\begin{equation}
-\partial_{x}\left[\left(v^{2}-\left(\mu\partial_{x}V\right)^{2}\right)\rho\right]-\alpha\rho\mu\partial_{x}V=0,\label{eq:zeroT equation}
\end{equation}
together with the boundary condition
\[
\rho(x)\underset{x\rightarrow-\infty}{\longrightarrow}\rho_{0}.
\]
\item \textbf{region (ii)}: Here we use the WKB-like Ansatz $\rho(x)=C_{D}(x)e^{-\frac{\varphi(x)}{D}}$ in Eq.\:(\ref{eq:close density}). The expression for $\varphi(x)$ is identical to that obtained using the methods of the main text. Note that also here to leading order $J=0$.
\item \textbf{region (iii)}: As in region (i) the contribution of diffusion terms $\propto D$ to the dynamics can be neglected. However, since the density of particles is now very low, the current $J$ is no longer negligible and one has to be accounted for it. Therefore, here we solve
\begin{equation}
-\partial_{x}\left[\left(v^{2}-\left(\mu\partial_{x}V\right)^{2}\right)\rho\right]-\alpha\rho\mu\partial_{x}V=\left(\alpha-\mu\partial_{x}^{2}V\right)J,\label{eq:zeroT transition}
\end{equation}
with the absorbing boundary condition $\rho(C_{3})=0$.
\end{enumerate}
The solutions found separately in regions (i), (ii) and (iii) have
to match together at the two points $C_{1}$ and $C_{2}$. To do this we have to calculate the structure of the solution near the the two points $C_{1}$ and $C_{2}$. These are given, as we detail below, by boundary layers of size $\sqrt{D}$ which can be matched to the solutions in the different regions.

\subsection{Solutions }
We next carry out the calculation outlined above in detail.

\subsubsection{Region (i)\label{subsec:region-1}}

The explicit solution of Eq.\:(\ref{eq:zeroT equation}) is 
\begin{equation}
\rho(x)=\frac{\rho_{0}v^{2}}{v^{2}-\left(\mu\partial_{x}V\right)^{2}}e^{-\alpha\int_{-\infty}^{x}\frac{\mu\partial_{y}V}{v^{2}-\left(\mu\partial_{y}V\right)^{2}}{\rm d}y}.\label{eq:zeroT solution1}
\end{equation}
In order to match this solution we have to understand how it behaves near $C_1$. To do this we make the change of variable $x\leftarrow x-C_{1}$. The force can be expanded according to
\[
\mu\partial_{x}V=v+k_{1}x+O(x^{2}).
\]
The equivalent of the integral in the exponential of Eq.\:(\ref{eq:zeroT solution1})
is
\[
\int_{-\infty}^{x}\frac{\mu\partial_{y}V}{v^{2}-\left(\mu\partial_{y}V\right)^{2}}{\rm d}y\underset{x\rightarrow0}{=}-\frac{1}{2k_{1}}\log\left(\frac{k_{1}\left|x\right|}{v}\right)+\gamma_1+O(\left|x\right|),
\]
where $\gamma_1$ is a finite constant that depends explicitly on the
potential through the relation
\begin{align*}
\gamma_1 & =\underset{x\rightarrow0}{\lim}\left\{ \int_{-\infty}^{x}\frac{\mu\partial_{y}V}{v^{2}-\left(\mu\partial_{y}V\right)^{2}}{\rm d}y+\frac{1}{2k_{1}}\log\left(\frac{k_{1}\left|x\right|}{v}\right)\right\} \\
 & =\mathfrak{F}\int_{-\infty}^{0}\frac{\mu\partial_{y}V}{v^{2}-\left(\mu\partial_{y}V\right)^{2}}{\rm d}y,
\end{align*}
where the last equality defines, as in the main text, the finite part of the diverging integral.
Note that there is some arbitrariness in the definition of the
finite part. Any function of the form $\frac{1}{2k_{1}}\log\left(\frac{\left|x\right|}{L}\right)$,
where $L$ is some arbitrary length scale, could be removed from the
integral to define the finite part. The above choice $L=\frac{v}{k_{1}}$
has been used in order to make the final expression for the mean escape
time more compact.

Restoring the original coordinate $x$, we therefore find
\begin{equation}
\rho(x)\underset{x\rightarrow C_{1}}{\sim}\frac{\rho_{0}}{2\left(\frac{k_{1}\left|x-C_{1}\right|}{v}\right)^{1-\alpha/2k_{1}}}e^{-\alpha\mathfrak{F} \int_{-\infty}^{C_1}\frac{\mu\partial_{y}V}{v^{2}-\left(\mu\partial_{y}V\right)^{2}}{\rm d}y}.\label{eq:left equivalent x1}
\end{equation}
The solution has two different behaviors depending on the value of
the second derivative $k_{1}=\mu\left.\partial_{y}^{2}V\right|_{x=C_{1}}$.
The density diverges at the critical point $x=C_1$ if $k_{1}>\frac{\alpha}{2}$,
and vanishes if $k_{1}<\frac{\alpha}{2}$. Since $k_{1}>0$ the diverging solution remains integrable
at $C_{1}$.
% This is analogous to the behavior already found numerically in {\bf CITE} and analytically in {\bf CITE}. 
We comment that it is straightforward to see that 
\[
\frac{P_{-}}{P_{+}}=\frac{v-\mu\partial_{x}V}{v+\mu\partial_{x}V}\underset{x\rightarrow C_{1}}{\sim}\frac{k_{1}}{2v}\left|x-C_{1}\right|.
\]
This implies that only right moving particles reach $C_1$.

\subsubsection{Region (ii)\label{subsec:region-2}}

In region (ii), we use the WKB-like Ansatz
\[
\rho(x)=C_{D}(x)e^{-\frac{\varphi(x)}{D}},
\]
where the large deviation pre-factor function can be expanded in powers of $D$
as
\[
C_{D}=C_{D}^{0}+DC_{D}^{1}+D^{2}C_{D}^{2}+...
\]
To leading order it is easy to check that, as expected, this reproduces the Eq.\:(10) of the main text for $\varphi(x)$. Using this solution with the expansion of the pre-factor we find to next order
\begin{equation}
\partial_{x}\varphi\,\partial_{x}C_{D}^{0}+\frac{\alpha}{2}C_{D}^{0}=0.\label{eq:prefactor}
\end{equation}
whose solution is 
\begin{equation}
C_{D}^{0}(x)=\overline{C}_{D}^{0}e^{-\frac{\alpha}{2}\int_{x_{0}}^{x}\frac{{\rm d}y}{\partial_{y}\phi}}.\label{eq:prefactor solution}
\end{equation}
with $x_{0}$ an arbitrary point between $C_{1}$ and $C_{2}$.

Again to match this solution we have to consider its behavior 
close to the two critical points $C_{1}$ and $C_{2}$. 
To this end, we make the  change of variable $x\leftarrow x-C_{1}$ and study the
behavior of $C_{D}^{0}(x)$ close to $C_{1}$. Close to $x=0$, we use the expansion
\[
\partial_{x}\varphi=k_{1}x+O(x^{2}),
\]
which shows that the integral in\:(\ref{eq:prefactor solution}) can
be expanded around $x=0$ as
\[
\int_{x_{0}}^{x}\frac{{\rm d}y}{\partial_{y}\varphi}=\frac{1}{k_{1}}\log\left(\frac{k_{1}x}{v}\right)-\gamma_2+O(x),
\]
where $\gamma_2$ is a finite constant, and we used Eq.\:(10) of the main text. Using the same notations as
in section\:\ref{subsec:region-1}, we have
\begin{align*}
\gamma_2 & =\underset{x\rightarrow0}{\lim}\left\{ \int_{x}^{x_{0}}\frac{{\rm d}y}{\partial_{y}\varphi}+\frac{1}{k_{1}}\log\left(\frac{k_{1}x}{v}\right)\right\} \\
 & =\mathfrak{F} \int_{0}^{x_{0}}\frac{{\rm d}y}{\partial_{y}\varphi} .
\end{align*}
Coming back to the original variable $x$, this gives
\[
C_{D}^{0}(x)\underset{x\rightarrow C_{1}}{\sim}\frac{\overline{C}_{D}^{0}}{\left(\frac{k_{1}\left|x-C_{1}\right|}{v}\right)^{\alpha/2k_{1}}}e^{\frac{\alpha}{2}\mathfrak{F} \int_{C_{1}}^{x_{0}}\frac{{\rm d}y}{\partial_{y}\varphi} }.
\]
The same line of arguments, gives the equivalent of the pre-factor close
to $C_{2}$ as
\[
C_{D}^{0}(x)\underset{x\rightarrow C_{2}}{\sim}\overline{C}_{D}^{0}\left(\frac{\left|k_{2}\right|\left|x-C_{2}\right|}{v}\right)^{\frac{\alpha}{2\left|k_{2}\right|}}e^{-\frac{\alpha}{2}\mathfrak{F}\int_{x_{0}}^{C_{2}}\frac{{\rm d}y}{\partial_{y}\varphi} },
\]
where $k_{2}=\mu\partial_{x}^{2}V(C_{2})$ is the (negative) second derivative
of the potential, and $\mathfrak{F} \int_{x_{0}}^{C_{2}}\frac{{\rm d}y}{\partial_{y}\varphi} $
is the finite part defined as 
\[
\mathfrak{F}\int_{x_{0}}^{C_{2}}\frac{{\rm d}y}{\partial_{y}\varphi} =\underset{x\rightarrow C_{2}}{\lim}\left\{ \int_{x_{0}}^{C_{2}}\frac{{\rm d}y}{\partial_{y}\varphi}+\frac{1}{\left|k_{2}\right|}\log\left(\frac{\left|k_{2}\right|\left|x-C_{2}\right|}{v}\right)\right\} \;.
\]
In what follows to match this solution with the other regions we note that the above results imply that close to $C_1$
\begin{equation}
\rho(x)\underset{x\rightarrow C_{1}}{\sim}\frac{\overline{C}_{D}^{0}}{\left(\frac{k_{1}\left|x-C_{1}\right|}{v}\right)^{\alpha/2k_{1}}}e^{\frac{\alpha}{2}\mathfrak{F} \int_{C_{1}}^{x_{0}}\frac{{\rm d}y}{\partial_{y}\varphi} }e^{-\frac{k_{1}\left|x-C_{1}\right|^{2}}{2D}},\label{eq:right equivalent x1}
\end{equation}
and close to $C_{2}$
\begin{equation}
\rho(x)\underset{x\rightarrow C_{2}}{\sim}\overline{C}_{D}^{0}\left(\frac{\left|k_{2}\right|\left|x-C_{2}\right|}{v}\right)^{\frac{\alpha}{2\left|k_{2}\right|}}e^{-\frac{\alpha}{2}\mathfrak{F}\int_{x_{0}}^{C_{2}}\frac{{\rm d}y}{\partial_{y}\phi} }e^{-\frac{\varphi(C_{2})}{D}-\frac{k_{2}\left|x-C_{2}\right|^{2}}{2D}}.\label{eq:left equivalent x2}
\end{equation}
These specify the boundary layers at the edges of region (ii). Their typical extension is $\sqrt{\frac{D}{k_{1}}}$
and $\sqrt{\frac{D}{\left|k_{2}\right|}}$ at $C_1$ and $C_2$ respectively.

\subsubsection{Region (iii) }

Eq.\:(\ref{eq:zeroT transition}) can be solved to give
\[
\rho(x)=\frac{J}{v^{2}-\left(\mu\partial_{x}V\right)^{2}}\int_{x}^{C_{3}}\left(\alpha-\mu\partial_{y}^{2}V\right)e^{\alpha\int_{x}^{y}\frac{\mu\partial_{z}V}{v^{2}-\left(\mu\partial_{z}V\right)^{2}}{\rm d}z}{\rm d}y.
\]
 Note that this expression is well defined, because $e^{\alpha\int_{x}^{y}\frac{\mu\partial_{z}V}{v^{2}-\left(\mu\partial_{z}V\right)^{2}}{\rm d}z}$
is integrable close to $C_{3}$. Using this we find that near $C_2$ the solution can be written as
\begin{equation}
\rho(x)\underset{x\rightarrow C_{2}}{\sim}\frac{J}{2v^{2}\left(\frac{\left|k_{2}\right|\left|x-C_{2}\right|}{v}\right)^{1+\frac{\alpha}{2|k_{2}|}}}\int_{C_{2}}^{C_{3}}\left(\alpha-\mu\partial_{y}^{2}V\right)e^{\alpha\mathfrak{F}\int_{C_{2}}^{y}\frac{\mu\partial_{z}V}{v^{2}-\left(\mu\partial_{z}V\right)^{2}}{\rm d}z}{\rm d}y,\label{eq:right equivalent x2}
\end{equation}
where again, the notation $\mathfrak{F}$ means 
\[
\mathfrak{F} \int_{C_{2}}^{y}\frac{\mu\partial_{z}V}{v^{2}-\left(\mu\partial_{z}V\right)^{2}}{\rm d}z =\underset{x\rightarrow C_{2}}{\lim}\left\{ \int_{x}^{y}\frac{\mu\partial_{z}V}{v^{2}-\left(\mu\partial_{z}V\right)^{2}}{\rm d}z+\frac{1}{2|k_{2}|}\log\left(\frac{\left|k_{2}\right|\left|x-C_{2}\right|}{v}\right)\right\} .
\]

\subsubsection{Matching at the boundary layers}

We now have to match all the solutions\:(\ref{eq:left equivalent x1},\ref{eq:right equivalent x1},\ref{eq:left equivalent x2},\ref{eq:right equivalent x2})
at the two critical points $C_{1}$ and $C_{2}$. 
To do this we need to solve the Fokker-Planck equation in the boundary layers around $C_1$ and $C_2$. 
To this end we define the variables $x-C_i=\sqrt{\frac{D}{|k_i|}}y_i$. Using this in Eq.\:(\ref{eq:close density}) we obtain to zeroth order in $D$ 
\begin{equation}
\frac{1}{{\rm sgn}(k_i)}\partial_{y_i}^{2}\rho_{k_i}+y\partial_{y_i}\rho_{k_i}+\left(1-\frac{\alpha}{2k_i}\right)\rho_{k_i}=0,\label{eq:BL-eq}
\end{equation}
where ${\rm sgn}(k)=\pm1$ denotes the sign of $k_i$. The solutions of this equation for large positive or negative values of $y_i$ have to be matched with the solutions in the different region. The solution for $i=1$ satisfies
\begin{equation}
\rho_{k_1}(y_1)\sim\begin{cases}
\frac{A_{D_1}}{{|y_1|}^{1-\frac{\alpha}{2k_1}}} & \text{when }y_1\rightarrow-\infty,\\
\frac{\Gamma\left(\frac{\alpha}{2k_1}\right)}{\sqrt{2\pi}}\frac{A_{D_1}}{{y_1}^{\alpha/2k_1}}e^{-\frac{{y_1}^{2}}{2}} & \text{when }y_1\rightarrow+\infty,
\end{cases}\label{eq:BL}
\end{equation}
and for $i=2$
\begin{equation}
\rho_{k_2}(y_2)\sim\begin{cases}
\frac{A_{D_2}}{{|y_2|}^{\alpha/2k_2}}e^{\frac{{y_2}^{2}}{2}}& \text{when }y_2\rightarrow-\infty,\\
\frac{\Gamma\left(1-\frac{\alpha}{2k_2}\right)}{\sqrt{2\pi}}\frac{A_{D_2}}{{y_2}^{1-\frac{\alpha}{2k_2}}} & \text{when }y_2\rightarrow+\infty,
\end{cases}\label{eq:BL}
\end{equation}
where $A_{D_1}$ and $A_{D_2}$ are two undetermined constant. By matching the asymptotic behavior\:(\ref{eq:BL}) of the boundary layer solution with the behavior of the solutions\:(\ref{eq:left equivalent x1},\ref{eq:right equivalent x1},\ref{eq:left equivalent x2},\ref{eq:right equivalent x2}) in the different regions close to $C_1$ and $C_2$ one finds after a lengthy calculations Eq.\:(12) of the main text. 

\subsection{Mean escape time from a metastable well  \label{sec_metastable}}
We now generalize our result to the mean escape time from a metastable
well. We introduce the critical point $C_{0}$ on the left of $C_{1}$
such that $\mu\partial_{x}V(C_{0})=-v$. The metastable well is represented in Fig.\:\ref{fig:metastable}.
\begin{figure}
\includegraphics[height=7cm]{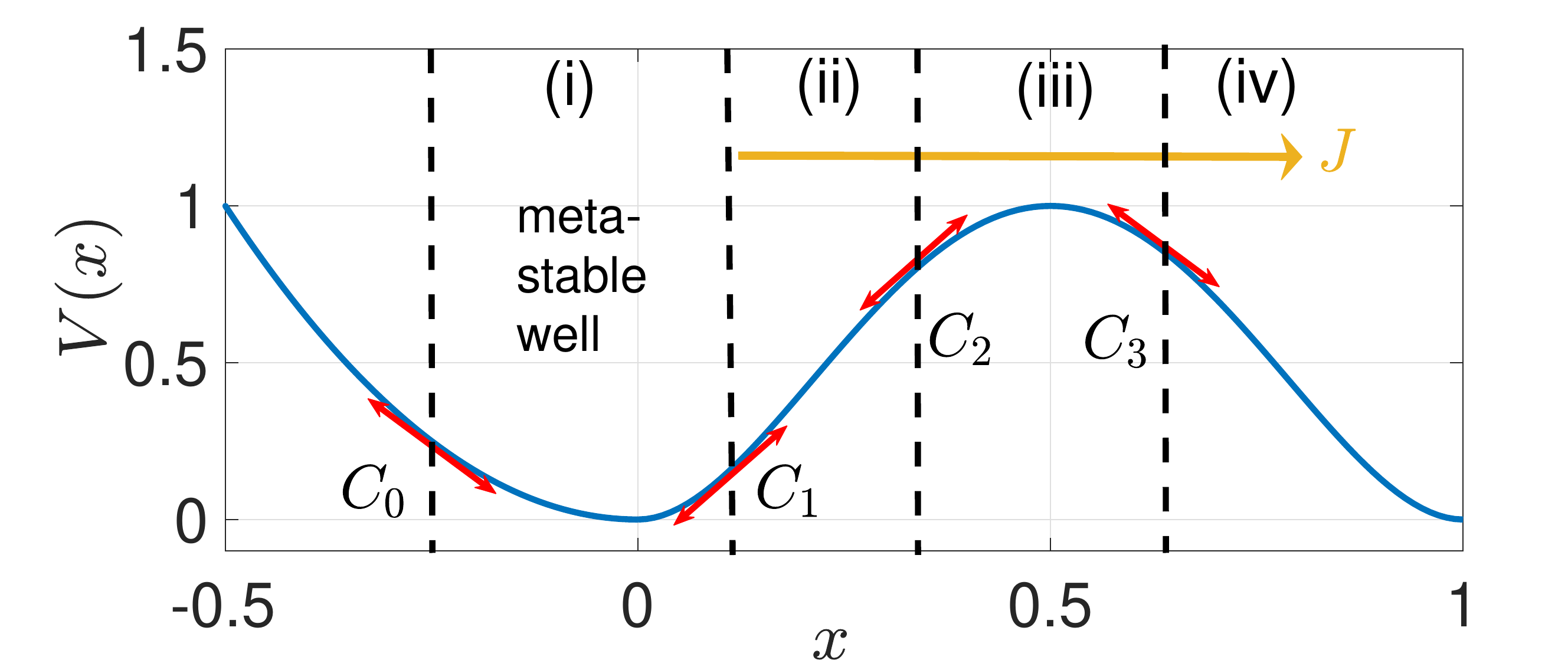}
\caption{Schematic representation of the active escape problem from a metastable well. At $D=0$, active particles are trapped between $C_0$ and $C_1$. When $D>0$, rare escape can occur through the right barrier, by crossing region (ii). \label{fig:metastable}}
\end{figure}
 According to expression\:(\ref{eq:zeroT solution1}),
the zero-fluctuations solution in region (i) writes 
\[
\rho(x)=\frac{Nv^{2}}{v^{2}-\left(\mu\partial_{x}V\right)^{2}}e^{-\alpha\int_{x_{b}}^{x}\frac{\mu\partial_{y}V}{v^{2}-\left(\mu\partial_{y}V\right)^{2}}{\rm d}y},
\]
where $x_{b}$ is some arbitrary point between $C_{0}$ and $C_{1}$, and $N$
is a constant, given by the normalization constrain $\int_{C_{0}}^{C_{1}}\rho(x){\rm d}x=1.$
We find
\[
N=\frac{1}{\int_{C_{0}}^{C_{1}}\frac{v^{2}}{v^{2}-\left(\mu\partial_{x}V\right)^{2}}e^{-\alpha\int_{x_{b}}^{x}\frac{\mu\partial_{z}V}{v^{2}-\left(\mu\partial_{z}V\right)^{2}}{\rm d}z}{\rm d}x}.
\]
 The mean escape time is then simply given by Eq.\:(12) replacing
the term
\[
\rho_{0}e^{-\alpha\mathfrak{F} \int_{-\infty}^{C_{1}}\frac{\mu\partial_{y}V}{v^{2}-\left(\mu\partial_{y}V\right)^{2}}{\rm d}y }
\]
by
\[
\frac{e^{-\alpha\mathfrak{F} \int_{x_{b}}^{C_{1}}\frac{\mu\partial_{y}V}{v^{2}-\left(\mu\partial_{y}V\right)^{2}}{\rm d}y }}{\int_{C_{0}}^{C_{1}}\frac{v^{2}}{v^{2}-\left(\mu\partial_{x}V\right)^{2}}e^{-\alpha\int_{x_{b}}^{x}\frac{\mu\partial_{z}V}{v^{2}-\left(\mu\partial_{z}V\right)^{2}}{\rm d}z}{\rm d}x}
\]
which can be equivalently written as
\[
\frac{1}{\int_{C_{0}}^{C_{1}}\frac{v^{2}}{v^{2}-\left(\mu\partial_{x}V\right)^{2}}e^{\alpha\mathfrak{F} \int_{x}^{C_{1}}\frac{\mu\partial_{z}V}{v^{2}-\left(\mu\partial_{z}V\right)^{2}}{\rm d}z }{\rm d}x}.
\]
We obtain the formula
\begin{eqnarray}
\left\langle \tau\right\rangle =\frac{2\pi}{\varGamma\left(\frac{\alpha}{2k_{1}}\right)\varGamma\left(1-\frac{\alpha}{2k_{2}}\right)}\frac{\left(\frac{\sqrt{Dk_{1}}}{v}\right)^{1-\frac{\alpha}{k_{1}}}}{\left(\frac{\sqrt{D\left|k_{2}\right|}}{v}\right)^{1-\frac{\alpha}{k_{2}}}}\int_{C_{0}}^{C_{1}}\frac{e^{\alpha\mathfrak{F} \int_{y}^{C_{1}}\frac{\mu\partial_{z}V}{v^{2}-\left(\mu\partial_{z}V\right)^{2}}{\rm d}z }}{v^{2}-\left(\mu\partial_{y}V\right)^{2}}{\rm d}y...\nonumber\\
...\times \int_{C_{2}}^{C_{3}}\left(\alpha-\mu\partial_{y}^{2}V\right)e^{\alpha\mathfrak{F} \int_{C_{2}}^{y}\frac{\mu\partial_{z}V}{v^{2}-\left(\mu\partial_{z}V\right)^{2}}{\rm d}z }{\rm d}y\,e^{\frac{\alpha}{2}\mathfrak{F} \int_{C_{1}}^{C_{2}}\frac{{\rm d}y}{\partial_{y}\varphi}}e^{\frac{\phi}{D}}.
\label{eq:metastable}
\end{eqnarray}

\section{Escape from a two-dimensional elliptic potential}

This section presents the computation of the quasi-potential for the
active escape problem out of the two-dimensional elliptic barrier described in the main text. The
potential can be written as
\[
\mu V(\mathbf{x})=\frac{1}{2}\mathbf{x}^{T}A\mathbf{x}\text{ for }V(\mathbf{x})<V_{0},
\]
where $A=\begin{pmatrix}\mu\lambda_{m} & 0\\
0 & \mu\lambda_{M}
\end{pmatrix}$ is a symmetric matrix of the second derivatives of the potential.
We consider without loss of generality that $0<\lambda_{m}<\lambda_{M}$.

Using the results of the main text, the fluctuation paths between specified initial and final positions are minimizers of the action
\begin{equation}
\mathcal{A}[\mathbf{x}(t)]=\frac{1}{4}\int_{-\infty}^{0}\left(\left\Vert \dot{\mathbf{x}}+A\mathbf{x}\right\Vert -v\right)^{2}{\rm d}t.\label{eq:action}
\end{equation}
To compute the fluctuation paths, we solve the Euler-Lagrange equation. As will become clear, it is useful to consider the momentum
\begin{equation}
\mathbf{p}(t)=\frac{\partial\mathcal{L}}{\partial\dot{\mathbf{x}}}=\frac{1}{2}\left(\dot{\mathbf{x}}+A\mathbf{x}\right)\left(1-\frac{v}{\left\Vert \dot{\mathbf{x}}+A\mathbf{x}\right\Vert }\right).\label{eq:impulsion}
\end{equation}
Interestingly for a quadratic potentials we find from Eq.\:(\ref{eq:action})
\[
\frac{\partial\mathcal{L}}{\partial\mathbf{x}}=A\frac{\partial\mathcal{L}}{\partial\dot{\mathbf{x}}}.
\]
Using this relation, the Euler-Lagrange equations then translate into
an equation for the momentum $\mathbf{p}$
\[
\dot{\mathbf{p}}=A\mathbf{p},
\]
whose solution is $\mathbf{p}(t)=e^{At}\mathbf{p}_{0}$. In the
present problem, $\mathbf{p}_{0}$ should be understood as the momentum
at the final position of the trajectory $\mathbf{x}(t=0)=\mathbf{x}_{\ff}$. The explicit
solution of $\mathbf{p}(t)$ together with Eq.\:(\ref{eq:impulsion}) gives the first order equation for ${\bf x}$
\begin{equation}
\frac{1}{2}\left(\dot{\mathbf{x}}+A\mathbf{x}\right)\left(1-\frac{v}{\left\Vert \dot{\mathbf{x}}+A\mathbf{x}\right\Vert }\right)=e^{At}  \mathbf{p}_{0}.\label{eq:eq_fluctuation}
\end{equation}

To solve Eq.\:(\ref{eq:eq_fluctuation}), we first take the norm
of both sides of the equality to get
\begin{equation}
\frac{1}{2}\left(\left\Vert \dot{\mathbf{x}}+A\mathbf{x}\right\Vert -v\right)=\left\Vert e^{At} \mathbf{p}_{0}\right\Vert ,\label{eq:norm impulsion}
\end{equation}
with the implicit assumption that the instanton path satisfies the condition $\left\Vert \dot{\mathbf{x}}+A\mathbf{x}\right\Vert >v$. Using (\ref{eq:norm impulsion}) in Eq.\:(\ref{eq:eq_fluctuation}),
we have
\begin{equation}
\dot{\mathbf{x}}+A\mathbf{x}=v\frac{e^{At} \mathbf{p}_{0}}{\left\Vert e^{At} \mathbf{p}_{0}\right\Vert }+2e^{At} \mathbf{p}_{0}.\label{eq:fluctuation}
\end{equation}
The boundary conditions for this equation are
\begin{equation}
\begin{cases}
\dot{\mathbf{x}}(t)\underset{t\rightarrow-\infty}{\longrightarrow} & 0\\
\mathbf{x}(t)\underset{t\rightarrow-\infty}{\longrightarrow} & \mathbf{x}_{1}\in\mathbf{C}_{1}
\end{cases}.\label{eq:BC}
\end{equation}
Eq.\:(\ref{eq:fluctuation}) together
with the constraints\:(\ref{eq:BC}) can only be satisfied if
$\mathbf{x}_{1}$ is an eigenvector of $A$. To see this, we expand of the right-hand side
of Eq.\:(\ref{eq:fluctuation}) in the limit $t\rightarrow-\infty$.
Because the two eigenvalues of $A$ satisfy $\lambda_m<\lambda_M$, we have $\left\Vert e^{At} \mathbf{p}_{0}\right\Vert \underset{t\rightarrow-\infty}{\sim} \ p_0^x e^{\lambda_m t}$ where  
$(p_{0}^{x},p_{0}^{y})$ are the two components of $\mathbf{p}_{0}$. We further have $ e^{At} \mathbf{p}_{0}=p_0^x e^{\lambda_m t}\mathbf{e}_{x}+p_0^y e^{\lambda_M t}\mathbf{e}_{y}$.
For $p_{0}^{x}\neq0$, the first term in Eq.\:(\ref{eq:fluctuation})  thus gives 
\[
\dot{\mathbf{x}}+A\mathbf{x}\underset{t\rightarrow-\infty}{\longrightarrow}{\rm sgn}(p_{0}^{x})v\mathbf{e}_{x},
\]
which proves, using $\dot{\mathbf{x}}(t)\underset{t\rightarrow-\infty}{\longrightarrow}0$,
that $\mathbf{x_{1}}=\left(\pm\frac{v}{\mu\lambda_{m}},0\right)$. When $p_0^x=0$ we find $\mathbf{x_{1}}={\bf x}_1^y=(0,\pm\frac{v}{\mu\lambda_{M}})$. This has a simple geometric interpretation. Generically $\mathbf{x}_{1}$, sitting on the $x$-axis, is a local extremum of $V(\mathbf{x})$ on the curve ${\bf C}_1$ (see Fig. (4) of the main text). With the exception of fluctuation paths which end on the $y$-axis all the paths start at one of the two local maxima located at $\mathbf{x_{1}}=\left(\pm\frac{v}{\mu\lambda_{m}},0\right)$.  This clearly minimizes the cost of the path. Fluctuation paths which end on the $y$-axis start at ${\bf x}_1^y$.
\\

We now turn to the full computation of the quasi-potential $\varphi(\mathbf{x}_{0})$,
where $\mathbf{x}_{0}$ is the final position of the fluctuation path.
The explicit expression of $\mathbf{x}_{0}$ can be computed from
the general solution of Eq.\:(\ref{eq:fluctuation})
\begin{equation}
\mathbf{x}_{0}(\mathbf{p}_{0})=v\int_{-\infty}^{0}\frac{e^{2At} \mathbf{p}_{0}}{\left\Vert e^{At} \mathbf{p}_{0}\right\Vert }{\rm d}t+A^{-1}\mathbf{p}_{0}.\label{eq:final position}
\end{equation}
Using Eq.\:(\ref{eq:norm impulsion}) in Eq.\:(\ref{eq:action}) and carrying out the integration in time we obtain the
large deviation rate function as a function of $\mathbf{p}_{0}$ 
\begin{equation}
\varphi(\mathbf{x}_{0}(\mathbf{p}_{0}))=\frac{1}{2}\mathbf{p}_{0}^{T}A^{-1}\mathbf{p}_{0}.\label{eq:LDFp0}
\end{equation}
Eqs.\:(\ref{eq:final position}) and\:(\ref{eq:LDFp0}) can both be solved
numerically to compute the quasi-potential $\varphi$ displayed in Fig. (4) of the main text. 

Besides, it is straightforward use Eq.\:(\ref{eq:final position}) to perform a small-fluctuations expansion around ${\bf p}_0$ in order to show that the action is minimal for paths moving only along the $x$-direction. Using this one can then easily compute the full expression for $\phi=\min\left\{ \varphi(\mathbf{x})|V(\mathbf{x})=V_{0}\right\} $.
We find
\[
\phi=\mu V_{0}\left(1-\sqrt{\frac{v^{2}}{2\mu^2\lambda_{m}V_{0}}}\right)^{2}.
\]
As expected, we recover the standard equilibrium result $\phi=\mu V_{0}$
when $v=0$.

\section{Supplementary information for the figures}
In this section, we provide the details about the potential $V(x)$ in each figure of the main text. We also describe the algorithm used for the numerics in Fig. 1, Fig. 3, and in the supplementary movie.

\subsection{Algorithm \label{sub:algo}}

For all the simulations presented in the main text, we adapted the Heun algorithm to simulate, for each individual particle, the following over-damped stochastic differential equation (see Eq. (1) in the main text)
\begin{equation}
\dot{\mathbf{x}}=v\mathbf{u}(\theta)-\nabla V(\mathbf{x})+\sqrt{2D}\boldsymbol{\xi}(t)\;.
\end{equation}
Here, ${\bf x}$ is the position of the particle and $v$ is its
self-propulsion speed. The orientation of the
particle ${\bf u}(\theta)$ evolves stochastically with a persistence
time $1/\alpha$. Note that compared to Eq. (1) of the main text, we have set $\mu=1$ everywhere.
$\boldsymbol{\xi}(t)$ is a vector of Gaussian white noise, such that
\begin{equation}
\xi_i(t)\xi_j(t')=\delta_{ij}\delta(t-t')\; .
\end{equation}
We discretized the time with time step $\delta t$, and updated the status of the particles according to
\begin{align}
\mathbf{x}^*&=\mathbf{x}(t)+\mathbf{v}\delta t-\nabla V(\mathbf{x}(t))\delta t+\sqrt{2D\delta t}\mathbf{W}_t\;, \\
\mathbf{x}(t+\delta t)&=\mathbf{x}(t)+\mathbf{v}\delta t-\frac{1}{2}[\nabla V(\mathbf{x}(t))+\nabla V(\mathbf{x}^*)]\delta t+\sqrt{2D\delta t}\mathbf{W}_t\;,
\end{align}
where $W_{t,i}\sim\mathcal{N}(0,1)$ is a normal distributed random number.

The reorientation of the particle is independent from its position. We thus sample the next tumbling time of each particle  from the exponential distribution $\alpha e^{-\alpha t}$. We split the time step where the tumbling happens into two smaller time steps: we first update the position of the particle until the time it tumbles, and then we uniformly randomly assign a new direction, and finish the remaining time.

We calculated the $C_1$, $C_2$, and $C_3$ numerically using false position method. The particles were considered to have escaped when their position reaches $C_3$.

\subsection{Figure 1\label{sub:fig1}}

The left potential barrier $V_{\LL}(x)$ in Fig. (1) is defined by

\begin{equation}
V_\LL(x)=\left\{\begin{array}{ll}
\frac{H_\LL\Delta_1(4+x/\ell_\LL)^2}{1-(1-\Delta_\LL)(4+x/\ell_\LL)^2}\;, & -4\ell_\LL\leq x <-3\ell_\LL\;, \\ 
2H_\LL-\frac{H_\LL\Delta_\LL(2+x/\ell_\LL)^2}{1-(1-\Delta_\LL)(2+x/\ell_\LL)^2}\;, & -3\ell_\LL\leq x <-\ell_\LL\;, \\ 
\frac{H_\LL\Delta_\LL(x/\ell_\LL)^2}{1-(1-\Delta_\LL)(x/\ell_\LL)^2}\;, & -\ell_\LL\leq x <0\;, \\ 
\end{array}\right.
\end{equation}
where the coefficients are defined through $\Delta_\LL=2/(\ell_\LL Z_\LL)$, $\ell_\LL=3$, $Z_\LL=1$ and $H_\LL=1.2$. The width of the barrier is $4\ell_\LL$, the height of the potential is $2H_\LL$ and the maximal slope is $H_\LL Z_\LL$.\\

The right potential barrier $V_{\RR}(x)$ in Fig. (1) is defined by
\begin{equation}
V_\RR(x)=\left\{\begin{array}{ll}
\frac{H_\RR\Delta_\RR(x/\ell_\RR)^2}{1-(1-\Delta_\RR)(x/\ell_\RR)^2}\;, & 0\leq x <\ell_\RR\;, \\ 
2H_\RR-\frac{H_\RR\Delta_\RR(2-x/\ell_\RR)^2}{1-(1-\Delta_\RR)(2-x/\ell_\RR)^2}\;, & \ell_\RR\leq x <3\ell_\RR\;, \\ 
\frac{H_\RR\Delta_\RR(4-x/\ell_\RR)^2}{1-(1-\Delta_\RR)(4-x/\ell_\RR)^2}\;, & 3\ell_\RR\leq x \leq 4\ell_\RR\;, \\ 
\end{array}\right.
\end{equation}
where the coefficients are defined through $\Delta_\RR=2/(\ell_\RR Z_\RR)$, $\ell_\RR=12$, $Z_\RR=2$ and $H_\RR=1$. The width of the barrier is $4\ell_\RR$, the height of the potential is $2H_\RR$ and the maximal slope is $H_\RR Z_\RR$.\\

The parameters for the Heun algorithm defined in section \ref{sub:algo} are listed in Table \ref{tab_fig_1_parameters}. We simulate each particles until it escapes from the barrier, that is, until it reaches $C_3$.

\begin{table}[htbp]
\begin{center}
\begin{tabular}{c|c|c|c|c|c}
\hline
\hline
$v$
&
$\delta t$
&
$D=0.08$
&
$D=0.0675$
&
$D=0.058$
&
$D=0.051$
\\
\hline
0.3
&
0.1
&
1117
&
237
&
&
\\
0.4
&
0.1
&
19329
&
2000
&
98
&
\\
0.45
&
0.1
&
&
&
&
396
\\
0.5
&
0.005
&
73319
&
5000
&
2100
&
2000
\\
0.6
&
0.005
&
70099
&
20000
&
20000
&
20000
\\
0.7
&
0.002
&
62639
&
40000
&
40000
&
40000
\\
0.8
&
0.002
&
100000
&
200000
&
40000
&
40000
\\
0.9
&
0.001
&
100000
&
200000
&
200000
&
200000
\\
1
&
0.001
&
100000
&
200000
&
200000
&
200000
\\
1.1
&
0.001
&
100000
&
200000
&
200000
&
200000
\\
\hline
\hline
\end{tabular}
\end{center}
\caption{\label{tab_fig_1_parameters} Time step sizes and numbers of samples of the simulations in Fig. (1) of the main text. The tumbling rate is $\alpha=1$.}
\end{table}

\subsection{Figure 3}
The potential in Fig. 3 is defined through
\begin{equation}
V(x)=\left\{\begin{array}{ll}
\infty\;, & x<0\;,\\ 
A\exp\left(C-\frac{C}{1-(x-B)^2/B^2}\right)\;, & 0\leq x\leq B\;,\\ 
0\;, & x>B\;,\\ 
\end{array}\right.
\end{equation}
where $A=1.5$, $B=1$, $C=2$. Those conditions correspond to  a  reflective boundary at $x=0$. We set $\alpha=1$, we use a time step $\delta t=0.001$, and $N_{\mathrm{samples}}=2\times 10^6$. The values of the particle's velocity are given by $v=1,\ 1.5,\ 2,\ 2.5$ respectively. We simulated each particles until it reaches $C_3$.

\subsection{Figure 4}
The functional dependence of the two barriers  is exactly the same as in Fig. 1 (see section \ref{sub:fig1}), with the parameters:\\
\textbf{Left barrier:} $\ell_\LL=3$, $Z_\LL=1$, $H_\LL=1$.\\
\textbf{Right barrier:} $\ell_\RR=8$, $Z_\RR=2$, $H_\RR=1$.\\
Contrary to the potential of Fig. (1), the left and right barriers have the same height here. The quasi-potential has thus the same value for the two barriers at $v=0$, and escape of passive particles is equally likely left and right, at least at the exponential level.

\subsection{Movie}

The potential $V(x)$ in the movie is built according to
\begin{equation}
V(x)=\left\{\begin{array}{ll}
3(x+3)^6-3\;, & x<-2\;,\\ 
1.5\exp\left(5-\frac{5}{1-(x+1)^2}\right)\;, & -2\leq x< 0\;,\\ 
\exp\left(0.1-\frac{0.1}{1-(x-1)^2}\right)\;, & 0\leq x< 2\;,\\ 
3(x-3)^6-3\;, & 2\leq x\;.\\ 
\end{array}\right.
\end{equation}
The two populations have $v=1.5$ (blue particles) and $v=3.3$ (red particles) respectively. Other parameters are $\alpha=1$, $D=0.06$, $\delta t=0.001$. $N_{\mathrm{samples}}=10^5$ particles are used to generate the histogram. The total time of the simulation is $t=20000$.

Active particles start in the metastable state around $x=0$, and escape left or right. When they escape, they are then trapped in the two deep wells located at $x=-3$ and $x=3$ respectively.

\end{widetext}

\end{document}